%% file: main.tex
\documentclass[sigconf]{acmart}
\AtBeginDocument{%
  \providecommand\BibTeX{{%
    \normalfont B\kern-0.5em{\scshape i\kern-0.25em b}\kern-0.8em\TeX}}}

\setcopyright{acmcopyright}
\copyrightyear{2018}
\acmYear{2018}
\acmDOI{10.1145/1122445.1122456}

\acmConference[Woodstock '18]{Woodstock '18: ACM Symposium on Neural
  Gaze Detection}{June 03--05, 2018}{Woodstock, NY}
\acmBooktitle{Woodstock '18: ACM Symposium on Neural Gaze Detection,
  June 03--05, 2018, Woodstock, NY}
\acmPrice{15.00}
\acmISBN{978-1-4503-XXXX-X/18/06}

\DeclareMathOperator{\arcosh}{arcosh}

\newcommand{\norm}[1]{\left\lVert#1\right\rVert}
\usepackage{caption}
\usepackage{subcaption}
\usepackage{multirow}
\begin{document}

%%
%% The "title" command has an optional parameter,
%% allowing the author to define a "short title" to be used in page headers.
\title{Fully Hyperbolic Graph Convolution Network for Recommendation}

\author[Liping Wang, Fenyu Hu, Shu Wu, and Liang Wang]{Liping Wang$^{1,2,}$*, Fenyu Hu$^{1, 2}$*, Shu Wu$^{1,2,\dagger}$, and Liang Wang$^{1,2}$}

\makeatletter
\def\authornotetext#1{
 \g@addto@macro\@authornotes{%
 \stepcounter{footnote}\footnotetext{#1}}%
}
\makeatother

\authornotetext{The first two authors made equal contribution to this work.}
%\authornotetext{The work is done during his internship at CRIPAC, CASIA.}
\authornotetext{To whom correspondence should be addressed.}

\affiliation{%
 \institution{\(^1\)Center for Research on Intelligent Perception and Computing, Institute of Automation, Chinese Academy of Sciences}
 \institution{\(^2\)School of Artificial Intelligence, University of Chinese Academy of Sciences}
 \country{}
}

\email{wangliping2019@ia.ac.cn,  fenyu.hu@cripac.ia.ac.cn}
\email{{shu.wu, wangliang}@nlpr.ia.ac.cn}

\def\authors{Liping Wang, Fenyu Hu, Shu Wu, and Liang Wang}
\renewcommand{\shortauthors}{Trovato and Tobin, et al.}

%%
%% The abstract is a short summary of the work to be presented in the
%% article.
\begin{abstract}
Recently, Graph Convolution Network (GCN) based methods have achieved outstanding performance for recommendation.
These methods embed users and items in Euclidean space, and perform graph convolution on user-item interaction graphs. 
However, real-world datasets usually exhibit tree-like hierarchical structures, 
%which makes Euclidean less suitable to embed users/items and conduct graph convolution.
which make Euclidean space less effective in capturing user-item relationship.
In contrast, hyperbolic space, as a continuous analogue of a tree-graph, 
provides a promising alternative.
In this paper, we propose a fully hyperbolic GCN model for recommendation, where all operations are performed in hyperbolic space. 
Utilizing the advantage of hyperbolic space, our method is able to embed users/items with less distortion and capture user-item interaction relationship more accurately.
Extensive experiments on public benchmark datasets show that our method outperforms both Euclidean and hyperbolic counterparts and requires far lower embedding dimensionality to achieve comparable performance.
\end{abstract}

%%
%% The code below is generated by the tool at http://dl.acm.org/ccs.cfm.
%% Please copy and paste the code instead of the example below.
%%
\begin{CCSXML}
<ccs2012>
 <concept>
  <concept_id>10010520.10010553.10010562</concept_id>
  <concept_desc>Computer systems organization~Embedded systems</concept_desc>
  <concept_significance>500</concept_significance>
 </concept>
 <concept>
  <concept_id>10010520.10010575.10010755</concept_id>
  <concept_desc>Computer systems organization~Redundancy</concept_desc>
  <concept_significance>300</concept_significance>
 </concept>
 <concept>
  <concept_id>10010520.10010553.10010554</concept_id>
  <concept_desc>Computer systems organization~Robotics</concept_desc>
  <concept_significance>100</concept_significance>
 </concept>
 <concept>
  <concept_id>10003033.10003083.10003095</concept_id>
  <concept_desc>Networks~Network reliability</concept_desc>
  <concept_significance>100</concept_significance>
 </concept>
</ccs2012>
\end{CCSXML}

\ccsdesc[500]{Computer systems organization~Embedded systems}
\ccsdesc[300]{Computer systems organization~Redundancy}
\ccsdesc{Computer systems organization~Robotics}
\ccsdesc[100]{Networks~Network reliability}

%%
%% Keywords. The author(s) should pick words that accurately describe
%% the work being presented. Separate the keywords with commas.
\keywords{recommendation system, graph neural networks, hyperbolic space}

%%
%% This command processes the author and affiliation and title
%% information and builds the first part of the formatted document.
\maketitle

\input{intro}
\input{pre}

\input{method}

\input{experiment}

\input{conclusion}

\newpage
\bibliographystyle{ACM-Reference-Format}
\bibliography{base}
\end{document}

%% file: intro.tex
\section{Introduction}

%\begin{figure}
%\centering
%\includegraphics[width=\linewidth]{figs/data_ori.pdf}	
%\caption{The number of multi-hop neighbors versus hops for ten sampled nodes from Amazon-CD dataset.}
%\label{fig:hop}
%\end{figure}

In information era, recommendation systems have been widely adopted to perform personalized information filtering \cite{covington2016deep, ying2018graph}.
%such as product recommendation in e-commerce website. 
%The core task is to predict whether a user would like to interact with an item, e.g. click and purchase.  
Even though there are many recommendation paradigms, collaborative filtering \cite{ebesu2018collaborative, ngcf} which generates recommendations by utilizing available historical interactions, remains a fundamental and challenging task.
%towards accurate personalized recommendation. 
The core idea behind collaborative filtering is to learn compact user/item embeddings 
 and infer a user's preference to an item according to the distance between their embeddings.
%Then, distances between embeddings of users and items  are used to infer users' preference. 
%Matrix factorization, the most popular CF method, embed users and items in a shared latent space and model a user's preference to an item as inner product between their embeddings. 
% \begin{figure}
% \centering
% \includegraphics{figs/ball.pdf}
% \caption{Digram of embedding users and items in Euclidean space}	
% \label{fig:embed_digram}
% \end{figure}

%\begin{figure}
%    \centering
%		\begin{subfigure}[t]{.45\linewidth}
%			\centering
%			\includegraphics[height=3cm]{figs/ball_left.pdf} %
%			\caption{\small}
%	\label{fig:single}
%		\end{subfigure}
%		\hfill
%		\begin{subfigure}[t]{.45\linewidth}
%			\centering
%			\includegraphics[height=3cm]{figs/ball_right.pdf} %
%			\caption{}
%			\label{fig:single}
%		\end{subfigure}
%		\hfill
%\caption{(a)User-item bi-partite graph, item colored according to the number of hops from $u_1$(without considering intermediate user nodes) (b)digram of embedding items in Euclidean space centering at $u_1$}
%\label{fig:ball}
%\end{figure}
From the perspective of graph, user-item interactions can be viewed as a bi-partite graph \cite{bollobas2013modern}, where nodes represent users/items and edges represent their interactions.
%between a user-item pair. 
 As a powerful tool of analyzing graph-structured data, Graph Neural Networks(GNNs) \cite{kipf2017semi, gat, graphsage} have recently demonstrated great success across various domains, including recommendation systems. 
 Employing multiple layers of neighborhood aggregation,  GNN-based methods \cite{ngcf, he2020lightgcn} have achieved  the state-of-the-art performance on diverse public benchmarks. 
 
 Although GNN-based methods have achieved outstanding performance, 
% we argue that 
 it might not be appropriate to adopt Euclidean space to embed users and items. 
% Analyzing several real-world datasets, we find that they exhibit tree-like hierarchical structure. 
In real-world scenarios, user-item bi-partite graphs usually exhibit tree-like hierarchical structures \cite{adcock2013tree},
%And the number of neighbors 
%As exemplified in Figure \ref{fig:hop}, in the user-item bi-partite graph of Amazon-CD dataset\cite{amazon_dataset}, 
in which the number of a node's neighbors grows exponentially with respect to the number of hops.
%the number of a user's high-order neighbors grows exponentially(linear in log-log plot) with respect to the number of hops.
% emb 与 hop 间的距离
% As illustrated in Figure \ref{fig:hop}, the number of a node's neighbors grows exponentially with respect to the number of hops(linear in log-log plot).
% To validate our thoughts, we analyze several real-world benchmark datasets carefully, and we find that 
% real-world CF dataset exhibits tree-like hierarchical structure where the number of node's neighborhood nodes grows exponentially with respect to the number of hops. 
Ideally, neighbors of node $v$ should be embedded in the ball centering at $v$'s embedding, and the distance between embeddings should reflect the number of hops between nodes.
Nevertheless, in Euclidean space, the volume of a ball only grows polynomially as a function of radius. 
%As a result, distortion would be brought to the embeddings of users and items in Euclidean space.
%As a result, embedding users and items in hyperbolic space would generate distortion. 
%Hence, this kind of distortion in embeddings would make it difficult to infer users' preference according to the distance between embeddings.
Hence, embedding exponentially-growing number of neighbors into polynomially-growing size of volume would make distance between embeddings less accurate to reflect distance between nodes in the graph, and this is called distortion \cite{hgcn}.
%However, in recommendation scenario, distance between user/item embeddings is used to infer users' preference. 
%Then this kind of distortion would make 
This kind of distortion makes it difficult to infer a user's preference to a target item according to the distance between their embeddings.
%As a result, embedding users and items in Euclidean space would introduce distortion and makes it difficult to infer users' preference according to the distance between embeddings.
 
 % tbd, link the gap
% As a result, projecting users and items into Euclidean space would inevitably bring distortion according to theoretical analysis\cite{linial1995geometry}. 
% To be more specific, the distortion comes from two sources. 
%For items, embedding are separated by a certain distance according to their similarity. The discrepancy between the volume of space and the number of neighbors may lead to item embeddings less discriminative. 
%For example, for an item $i$, compared to its first-order neighbors(users who have purchased item $i$), its second-order neighbors(items co-purchased by users ) should be further away from $i$ in embedding space. 
%However, those neighbors would be squeezed and less discriminative due to the inconsistency between available volume and the number of neighbors.  
%Similarly, users with less common preference would be squeezed closer 
%%due to the lag of volume growth in Euclidean space. 
%% due to the limited volume compared with 
% due to exponentially-growing number of neighbors.

In contrast, hyperbolic space \cite{hgcn} in which the volume of a ball grows exponentially with radius offers a promising alternative.
Compared with Euclidean space, hyperbolic space is more suitable for modeling user-item interaction graphs which exhibit strong tree-like hierarchical structures.
%And it is naturally suitable for embed users and items 
%And it is a natural choice to embed users and items in hyperbolic space due to more accurate 
Accordingly, it is a natural choice to conduct user/item embedding and graph convolution in hyperbolic space for recommendation. 
%And it is naturally suitable for modeling tree-like structure data\cite{hgnn, hgcn}. 
%Theoretically,  it can be viewed as a continuous analogue of tree graphs and can embed trees with arbitrarily low distortion\cite{krioukov2010hyperbolic}.
To the best of our knowledge, the only work 
%combining GNN and hyperbolic space 
adopting a similar idea 
is HGCF \cite{hgcf}.
%However, HGCF resorts to tangent space, which can be viewed as first-order approximation at an arbitrary point, to conduct graph convolution operation. 
%And this would bring 
However, resorting to tangent space to realize graph convolution  makes the performance of HGCF inferior
%since tangent space is only local first-order approximation.
for the two following reasons.
%\begin{itemize}
%	\item 
%	The core task of recommendation system is to learn accurate embedding for users and items according available user-item interactions. 
%	Nevertheless, 
On the one hand,
% in HGCF, user-item interaction relationship is approximated by tangent space
tangent space is only local linear approximation of hyperbolic space. 
%	Distortion caused by approximation may broadcast to  as message propagates between neighbors. 
%Error due to approximation spread to neighbors as message passes and aggregates, disturbing the learning of user/item embedding. 
During the process of message propagation, errors caused by approximation accumulate and spread to the whole graph.
As a result, influence from high-order neighbors cannot be captured accurately.
%		As a result, noisy and inaccurate embeddings of users and items would have an considerable negative impact on performance
%	\item
On the other hand, tangent space is actually Euclidean space according to its definition \cite{boothby1986introduction}. 
Hence, the advantage of hyperbolic space in modeling user-item interaction relationship with less distortion cannot be fully utilized.
%		which can cause inevitable errors for user/item embedding. 
%		Then, information from high-order neighbors may generate a negative impact.
%	\item Tangent space is actually Euclidean space, in which HGCF performs graph convolution. As a result, the advantage of hyperbolic space in modeling user-item interaction relationship with less distortion cannot be fully utilized.
	
%\end{itemize} 

%tbd, hgcn
 
% As Figure \ref{fig:embed_digram} shows, in embedding space, items lie in the neighborhood of their purchaser. But, forcing exponentially-growing item nodes into a polynomially-growing volume space would make those items overlap and less discriminative. 
% 

%Notably, all these methods are conducted in Euclidean space. 
%Recently, there are some studies show that many real world dataset exhibits typical characteristics of complex networks i.e. a high degree of clustering and scale free.
%And user-item interaction bi-partite graph can be seen as a generative model of complex network. 
%As a result, embedding user and item in Euclidean space would result distortion, hindering performance of recommendation system. 
%Hyperbolic space, as a continuous analogue of tree graphs, has recently demonstrated superiority modeling tree-like hierarchical graphs. 
To overcome the limitation of Euclidean space and 
%fully unleash advantages of hyperbolic space and GNN, 
obtain more accurate user/item embeddings,
we design a novel fully hyperbolic GCN framework specially for collaborative filtering. 
All operations are conducted in hyperbolic space, more specifically, in the Lorentz model, and we name it \textbf{L}orentz \textbf{G}raph \textbf{C}ollaborative \textbf{F}iltering (LGCF). 
%Naturally, this has motivated representation learning in hyperbolic space to model user-item graph structure more accurately. 
%Empirically, we found user-item bi-partite graph also display such characteristics. As Figure \ref{fig:user_degree} shows, the number of interactions follow a powerlaw distribution. 
%Even though there have been several hyperbolic embedding learning methods targeting at node classification or link prediction, such as HAN, HGNN and HGCN, direct adoption to recommendation system can hardly ensure desirable performance. 
 
%To the best of our knowledge, the only specially designed GNN-based method for recommendation is HGCF. However, resorting to tangent spaces to realize graph convolution  on hyperbolic manifold makes performance inferior since tangent space is only local first-order approximation. 
 
The main contributions of this work are summarized as follows:
\begin{itemize}
%	\item We are the first to point out that real-world 
%recommendation datasets
%exhibit tree-like structure and Euclidean space is inferior in modeling them. 
	\item We propose a fully hyperbolic graph convolution network for recommendation.
	\item We conduct extensive experiments on multiple public benchmark datasets, and  the results demonstrate the superiority of our method. 
\end{itemize}

%% file: pre.tex
\section{Preliminaries}
%In this section, we first set up the problem and provide mathematical notations for later use. After that, brief but essential background knowledge about hyperbolic space is provided.
%\subsection{Problem Formulation}

%In this section, we provide necessary background knowledge about hyperbolic space. And we offer detailed explanation about reasons why hyperbolic space is chosen and its specific advantages in modeling real-word dataset graphs. 

\textbf{Problem Formulation}. In this paper, the standard collaborative filtering set-up is considered.
Let $U=\{u_1, u_2, \dots, u_n\}$ be the set of $n$ users, and $I=\{i_1, i_2, \dots, i_m\}$ be the set of $m$ items. 
Historical interactions between users and items are represented as a binary matrix $B$, where $B_{i,j}=1$ if the $i$-th user interacts with the $j$-th item, otherwise $0$.
Given historical interactions $B$, the goal is to predict potential interactions. 

%\subsection{GNN-based Recommendation Methods}
\textbf{GNN-based Recommendation Methods}. GNN-based recommendation methods have received increasing attention for their ability to learn rich node representations.
In collaborative filtering setting, GCN has replaced matrix factorization \cite{bprmf} and shown leading performance.
In order to capture the influence from high-order neighbors, Wang et al. \cite{ngcf} proposed NGCF, a GCN framework performing on user-item interaction bi-partite graphs.   
Subsequently, He et al. \cite{he2020lightgcn} empirically found that two most common designs in GCNs --- feature transformation and nonlinear activation, contribute little to the performance of collaborative filtering. 
Hence, they proposed a simplified architecture -- LightGCN \cite{he2020lightgcn}.

%% file: method.tex
\section{Our Method}
%Motivated by characteristic advantages of hyperbolic space and huge success demonstrated by graph convolution networks recently. 
%In this section, overall design of our method and individual components 
%The former section demonstrates that tree-like structure appearing in real-world datasets which weaken the competitiveness of Euclidean space. In contrast, hyperbolic space, as a continuous analogue of tree graphs, provides a potential alternative.
%Driven by these findings, we set the goal of developing a competitive model which conduct all operations in hyperbolic space for recommendation.
% 

%In this section, we first set up the problem, then an overview of our method LGCF is provided. 
%In this section, we first provide an overview of LGCF.
%Then, procedures of each component are described in detail. 
%In the end, optimization process is offered for a better understanding.

\begin{figure*}
		\includegraphics[width=0.95\linewidth]{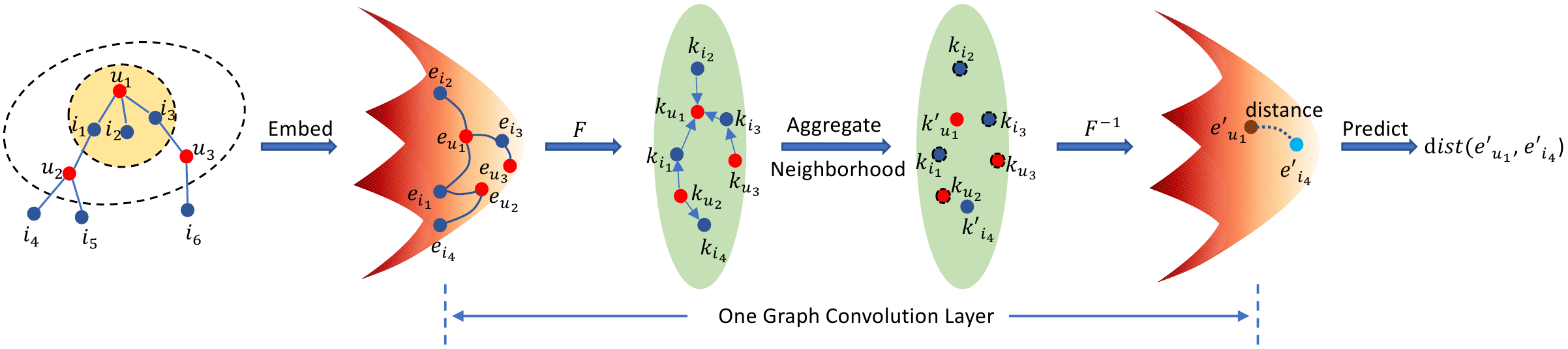}
		\caption{Illustration of LGCF. First, users and items are embedded into the Lorentz model of hyperbolic space. Then, multiple graph convolution layers (only one shown in the figure for simplicity) are adopted to aggregate information from neighbors. In each layer, embeddings are first mapped by $F$ to the Klein model in which graph convolution is performed. After that, $F^{-1}$ maps refined embeddings back to the Lorentz model. Finally, LGCF infers a user's preference to an item according to the distance between their embeddings.}
	\label{fig:model}
\end{figure*}

%In this section, we present the proposed LGCF model, the architecture of which is illustrated in  Figure \ref{fig:model}.
As illustrated in Figure \ref{fig:model}, 
there are three components in LGCF: (1) an embedding layer that provides and initializes user/item embeddings in hyperbolic space; (2) multiple graph convolution layers 
%(only one kept in the figure for simplicity) 
that propagate user/item embeddings over the graph; 
%(3) a prediction layer that outputs the distance between embeddings of a user-item pair as the probability of interaction.
and (3) a prediction layer that estimates a user's preference to an item by computing the distance between their embeddings.

%\subsection{Overview of LGCF}
%In LGCF, both users and items are embedded the
%Lorentz model of hyperbolic space.
%%hyperbolic space, more specifically, a Lorentz model. 
%Similar to existing GNN-based recommendation methods\cite{ngcf, he2020lightgcn}, stacking-layer manner is  adopted in LGCF. 
%And in each layer, neighborhood aggregation and nonlinear activation is conducted in sequence. 
%Even though feature transformation is the key to the success of modern neural networks, it is discarded in LGCF. 
%The reason is that: different from graphs where nodes bring rich attribute information, nodes in user-item interaction graphs contain no create semantics but one-hot IDs.
%In this case, feature transformation may provide no benefits, and could bring difficulties to training. 

\subsection{Embed Users/Items in Hyperbolic Space}
%In LGCF, both users and items are embedded in hyperbolic space, specifically, in the Lorentz model.
Existing GNN-based recommendation methods usually embed users and items in the same Euclidean space.
To accurately model user-item interaction relationship, we investigate the utilization of hyperbolic space.
There are several models of hyperbolic space, such as the Lorentz model, the Klein model and the Poincar\'e ball model.
In this paper, we choose the Lorentz model due to its simplicity and numerical stability.
Formally, the Lorentz model of $d$-dimensional hyperbolic space is defined as:
\begin{equation}
	\mathcal L = \{\boldsymbol x = [x_0, x_1, \cdots, x_d]\in \mathbb R^{d+1}:\left<\boldsymbol x, \boldsymbol x\right>_{\mathcal L} =-1, x_0 > 0 \},
\end{equation}
where $\left<\boldsymbol x, \boldsymbol y\right>_{\mathcal L}$ is Lorentz inner product and is defined as $\left<\boldsymbol x, \boldsymbol y\right>_{\mathcal L}  = -x_0y_0 + \sum_{i=1}^d x_iy_i$.
In addition, at an arbitrary point $\boldsymbol{x} \in \mathcal L$, hyperbolic space can be locally approximated by a linear Euclidean space. 
And this approximated Euclidean space is termed as tangent space  $\mathcal T_{\boldsymbol{x}}\mathcal L$ in which norm $\norm {\boldsymbol{x}}_{\mathcal L} = \left<\boldsymbol x, \boldsymbol y\right>$ is well defined.
In LGCF, both users and items are embedded in the same Lorentz model of hyperbolic space.
%To be more specific, Lorentz model is adopted, 
%in which both representation learning and graph convolution are conducted. 
%And they are refined 
%We refine the embeddings by propagating them on the user-item interaction graph.

%Notably, embeddings are the only parameters of LGCF and they are optimized in an end-to-end fashion.
%This leads to more accurate and effective embeddings since embedding refinement step inject collaborative signal into embeddings explicitly. 
%A key component to representation learning is embedding initialization, which has a huge impact on  the difficulty of optimization and the quality of final solution obtained. 

It is well known that random initialization can have a significant impact on optimization in training \cite{sutskever2013importance}.
A common practice in Euclidean space is Gaussian distribution initialization. 
Similarly, we design an initialization strategy for embeddings based on Wrapped Normal Distribution \cite{nagano2019wrapped} which is a generalization of Gaussian distribution to hyperbolic space.
\subsection{Graph Convolution Layer}
%In LGCF, embeddings of users and items are refined by performing graph convolution iteratively over the graph.
%Nevertheless, 
The basic idea of GCN-based recommendation models is learning representations for users and items by aggregating neighbors' information iteratively over the interaction graph. 
%To achieve this goal, graph convolution is performed iteratively.
%Inspired by this, in LGCF, we want to refine embeddings of users and items through graph convolution.
%Nevertheless, existing GCN models are defined in Euclidean space, and it is non-trivial to extend them to hyperbolic space.
%%This idea can be applied to recommendation by 
%To conduct graph convolution operation for recommendation in hyperbolic space, we design graph convolution layer for embeddings in hyperbolic space specifically.
In order to apply GCN for recommendation where users and items are embedded in hyperbolic space, we design graph convolution layers specially since na\"ive generalization will drive embeddings out of hyperbolic space.
Before that, we give a brief review of existing graph convolution layers in Euclidean space.

In Euclidean space, a graph convolution operation is composed of three steps: feature transformation, neighborhood aggregation and non-linear activation. 
Among them, feature transformation is performed through linear transformation. 
In LGCF, we discard the feature transformation operation for two reasons. 
%On the one hand, in recommendation scenario, because embeddings contains no rich semantics information, feature transformation 
On the one hand, different from attributed graphs (e.g., citation networks) where nodes bring rich feature information, nodes in user-item interaction graphs contain no semantics but one-hot IDs.
In this case, feature transformation may provide no benefits, and could bring difficulties to training.
%On the other hand, linear transformation itself is not in harmony with hyperbolic space and is almost impossible to extend to hyperbolic space.
%On the other hand, it is unnatural to extend linear transformation (matrix-vector multiplication) to hyperbolic space since it is not a vector space.
On the other hand, linear transformation (matrix-vector multiplication) is not well-defined in hyperbolic space since it is not a vector space. 

\textbf{Neighborhood Aggregation}.
Exising neighborhood aggregation of the $l$-th layer can be summarized as:
\begin{equation}
	\boldsymbol e^{(l)}_i = \sum_{j\in \hat{\mathcal N}(i)} w_{ij}\boldsymbol e_j^{(l-1)},
\end{equation}
%\begin{equation}
%	\boldsymbol{h}^{(i)} \gets \sum_{j\in hat N(i)} w_{ij} \boldsymbol{h}^{(j)}
%\end{equation}
%which can be viewed as a form of weighted mean.
in which $\boldsymbol e_j^{(l-1)}$ represents the $j$-th node's embedding and $\hat{\mathcal N}(i)$ denotes the set consisting of node $i$ and its neighborhood nodes.
%When it comes to hyperbolic space, a natural generalization of mean is Fr\'echet mean\cite{pennec2006intrinsic}. 
%However, it is difficult to bring Fr\'echet mean into application due to the missing of closed-form solution. 
%However, mean operation is 
A natural generalization of mean to hyperbolic space is Einstein midpoint \cite{ungar2008gyrovector}.
%which is defined as:
However, it is defined in the Klein model $\mathcal{K}=\left\{\boldsymbol x \in \mathbb{R}^{d}:\|\boldsymbol x\|<1\right\}$, while user/item embeddings lie in the Lorentz model.
Luckily, there are isometric maps between the Klein model and the Lorentz model defined as follows:
%\begin{align}
%	\begin{split}
%		f_{\mathcal L^d\rightarrow \mathcal K^d}(\boldsymbol{x}) &= \frac{[x_1, x_2, \dots x_d]}{x_0},\\
%		f_{\mathcal K^d\rightarrow \mathcal L^d}(\boldsymbol{k}) &= \frac{1}{1 - \norm{\boldsymbol{k}}^2}[1, \boldsymbol k],
% 	\end{split}
%\end{align}
\begin{align}
	\begin{split}
		F(\boldsymbol{x}) &= \frac{[x_1, x_2, \dots x_d]}{x_0},\\
		F^{-1}(\boldsymbol k) &= \frac{1}{1 - \norm{\boldsymbol{k}}^2}[1, \boldsymbol k],
 	\end{split}
\end{align}
in which $\boldsymbol{x} = [x_0, x_1, \dots, x_d]\in \mathcal L$ and $\boldsymbol{k} \in \mathcal K$.
%let $x\in \mathcal L^d$, then

Hence, we propose a neighborhood aggregation strategy utilizing the Klein model as an intermediate bridge.
Specifically, neighborhood aggregation can be divided into three steps. 
%Let $ \boldsymbol h_j$ be current embedding for node $j$.
First, current embeddings $\{\boldsymbol e_1, \boldsymbol e_2, \dots, \boldsymbol e_n\}$ are mapped from the Lorentz model to the Klein model by $\boldsymbol{k}_i=F(\boldsymbol e_i)$.
%\begin{equation}\label{eq:l2k}
%	f_{\mathcal L \rightarrow \mathcal K}\left ( \boldsymbol h_j\right )  = \frac{1 - \norm{\boldsymbol{h_j}}^2}{[1, \boldsymbol h_j]}. 
%\end{equation}
%\begin{equation}\label{eq:l2k}
%	\boldsymbol{k}_i = \frac{1 - \norm{\boldsymbol{e_i}}^2}{[1, \boldsymbol e_i]},
%\end{equation}
%in which $[1, \boldsymbol h_i]$ means appending an $1$ to the beginning of vector $\boldsymbol{h}_i$.
Then, neighborhood aggregation is conducted in the Klein model as follows:
\begin{align}
	\begin{split}
		\gamma_i =\frac{1}{\sqrt{1-\norm{\boldsymbol k_j}^2}},\\
		\boldsymbol k_i'  = \frac{\sum_{j\in \hat N(i)}\gamma_j \boldsymbol k_j}{\sum_{j\in \hat N(i)}\gamma_j}.\\
	\end{split}
\end{align}
Here, we obtain aggregated user/item embeddings 
%by aggregate neighborhood 
in the Klein model.
Last, aggregated embeddings $\boldsymbol k_i'$ are mapped back to the Lorentz model through $\boldsymbol z_i = F^{-1}(\boldsymbol k_i')$.

\textbf{Nonlinear Activation Layer}.
In Euclidean space, nonlinear activation has proven to be a key component in modern neural networks. 
However, direct adoption will drive the computation result out of hyperbolic space. 
To fix this problem, we design a calibration strategy following general activation.
%following normal nonlinear activation, 
Formally, let $\boldsymbol{x_i} = [x_0, x_1, \dots, x_d]=\sigma(\boldsymbol{z}_i)$ be the output of general activation function $\sigma$, e.g., ReLU. 
The first element of $\boldsymbol{x}$ is calibrated while the other elements remain unchanged to pull the activated embedding back to hyperbolic space:
\begin{equation}
	\boldsymbol e_i' = f_c(\boldsymbol{x}_i) = \left[\sqrt{1 + \sum_{j=1}^d x_j^2}, x_1,x_2, \dots, x_d\right].
\end{equation}
%Put together, calibrated nonlinear activation can be formulated as:
%$f_c(\sigma(x))$, 
%in which $\sigma$ is ReLU activation. 

\subsection{Prediction Layer}
After propagating with $L$ graph convolution layers, we obtain multiple representations for users and items. 
Representations generated in different layers emphasize messages passed through different connections, and they reflect users' preference or items' attributes from different perspectives. 
%Since these representations lie in hyperbolic space, we 
%Recommendation models operating in hyperbolic space usually employ the inner product between representations of a user and a target item to estimate 
Recommendation models operating in hyperbolic space usually estimate a user's preference to a target item according to the distance or similarity metric between their representations. 
In the Lorentz model, the generalization of a straight line in Euclidean space is geodesics which gives the shortest distance between two points 
$\boldsymbol{x}, \boldsymbol{y}$.
Formally, the geodesics distance between $\boldsymbol{x}$ and $\boldsymbol{y}$ is defined as:
\begin{equation}
\label{eq:geodesics}
d_{\mathcal L}(\boldsymbol{x}, \boldsymbol{y}) = \arcosh(-\left<\boldsymbol x, \boldsymbol y\right>_{\mathcal L}).
\end{equation}
Hence, in LGCF, we infer the preference of user $u$ to item $i$ based on the geodesics distance between their corresponding representations. 
Further, to utilize different semantics captured by different layers, we take representations learned by different layers into consideration simultaneously.
In summary, LGCF estimates the preference of user $u$ to target item $i$ as:
\begin{equation}
	y(u, i) = \frac{1}{\sum_{l=1}^{L} d_{\mathcal L}^2(\boldsymbol e_u^{(l)}, \boldsymbol e_i^{(l)})},
\end{equation}
in which $\boldsymbol e_u^{(l)}$ and $\boldsymbol e_i^{(l)}$ are representations generated by the $l$-th layer.
Even though there are multiple choices for layer aggregation, such as weighted average, max pooling, LSTM, etc., we find simple summation adopted here works well empirically.
%we adopt geodesics distance to measure the similarity between user-item representation pair. 

\subsection{Margin Ranking Loss}
Margin ranking loss \cite{tay2018latent} has been a competitive choice for distance-based recommendation systems. 
Since it encourages positive and negative user-item pairs to be separated up to a given margin. 
Once the difference between a negative user-item pair and a positive one is greater than the margin, these two user-item pairs will make no contribution to the loss.
In this way, hard pairs violating the margin are focused all the time, making optimization much easier. 
We extend margin ranking loss to hyperbolic space based on geodesics distance.
Given a sampled positive user-item pair $(u, i)$ and a negative one $(u, j)$, geodesics margin loss is defined as:
\begin{equation}
	\ell_g(u, i, j) = \sum_{l=1}^L\min \left(d_{\mathcal L}^2(\boldsymbol e_u^{(l)}, \boldsymbol e_j^{(l)})- d_{\mathcal L}^2(\boldsymbol e_u^{(l)}, \boldsymbol e_i^{(l)}) -m , 0 \right),
\end{equation}
where $m$ is a non-negative hyper-parameter.
%in which $d_{\mathcal L}$ is geodesics in Lorentz model defined in Eq \ref{eq:geodesics} and $\boldsymbol z_u, \boldsymbol z_i, \boldsymbol z_j$ are embeddings for user $u$ and item $i$ and $j$ respectively.
Note that representations obtained by different layers contribute to the loss simultaneously. 
%In addition to utilizing different-level semantics, 
%As suggested by 
This not only makes it possible to utilize different semantics captured by different layers, but also decreases the difficulty of optimization due to the residual connection \cite{resnet}.
%\subsection{Layer Combination}
%Previous work\cite{he2020lightgcn} has found that the embeddings would be over-smoothed with the increasing of the number of layers and embeddings at different layers capture different semantics. 
%Thus using only the last layer is problematic. 
%Instead, in LGCF, embeddings of different layers are combined together through individual loss\cite{ride_iclr21}, i.e., Geodesics Margin Losses computed according to embeddings of each layer are summed together:
%\begin{equation}
%	L(u, i, j) = \sum_{i=1}^l \ell_g(z_u^{(i)}, z_i^{(i)}, z_j^{(i)}).
%\end{equation}

\subsection{Optimization}
The only parameter of LGCF is the embedding matrix of users and items. 
These embeddings lie in the Lorentz model of hyperbolic space which is out of the range of common optimization algorithms such as SGD \cite{sgd} and Adam \cite{adam}. 
Hence, we employ RGSD \cite{rsgd}, a generation of SGD to hyperbolic space, which mimics SGD's behavior while taking into account the geometry of hyperbolic space. 
%Formally, the $t-$th of RSGD at point $\boldsymbol\theta^{(t)}$ can be split into three sub-steps: 
%\begin{enumerate}
%	\item Compute Euclidean gradient $\boldsymbol g_e^{(t)}$ like SGD.
%	\item Project Euclidean gradient to tangent space $\mathcal T_{\boldsymbol\theta} \mathcal L$ to obtain Hyperbolic gradient as:
%	\begin{equation}
%		\boldsymbol g_h^{(t)}=\boldsymbol g_e^{(t)} + \left <\boldsymbol\theta^{(t)}, \boldsymbol g^{(t)}\right >_{\mathcal L}\boldsymbol \theta^{(t)}
%	\end{equation}
%	\item Conduct gradient descent by mapping descent step $-\eta \boldsymbol g_h$ to hyperbolic space from tangent space through exponential map  $\exp_{\boldsymbol{\theta}}(\boldsymbol{x}) = \cosh(\norm{\boldsymbol x}_{\mathcal L})\boldsymbol{\theta}+ \sinh(\norm{\boldsymbol x}_{\mathcal L})\frac{\boldsymbol x}{\norm {\boldsymbol x}_{\mathcal L}}$ as:
%	\begin{equation}
%		\boldsymbol\theta^{(t+1)} = \exp_{\boldsymbol\theta}^{(t)}(-\eta \boldsymbol g_h^{(t)}),
%	\end{equation}
%\end{enumerate}

%% file: experiment.tex
\section{Experiments}
%We conduct extensive experiments on public benchmark and compare LGCF's result with 
\subsection{Set Up}
\begin{table}[]
\caption{Dataset statistics.}
\label{tbl:data}
\begin{tabular}{l|ccc}
\toprule
Dataset     & \#User & \#Item & \#Interactions \\
\midrule
Amazon-CD   & 22,947 & 18,395 & 422,301        \\
Amazon-Book & 52,406 & 41,264 & 1,861,118       \\
Yelp2020    & 91,174 & 45,063 & 1,940,014      \\
\bottomrule
\end{tabular}
\end{table}

\textbf{Datasets and Baselines}. 
Following HGCF \cite{hgcf}, we employ Amazon-CD \cite{amazon_dataset}, Amazon-Book \cite{amazon_dataset} and Yelp2020 \cite{yelp_dataset} datasets.
Dataset statistics are provided in Table \ref{tbl:data}. 
Each dataset is split into 80-20 train and test sets. 
%All methods are evaluated on test sets by Recall and NDCG metrics. 
Multiple competitive baseline methods from three categories are compared: BPRMF \cite{bprmf}, NGCF \cite{ngcf}, LightGCN \cite{he2020lightgcn}, HVAE \cite{hvae}, HGCF \cite{hgcf}. Among them, BPRMF optimizes matrix factorization by Bayesian personalize ranking (BPR) loss \cite{bprmf}. 
NGCF and LightGCN employ GNN in Euclidean space. 
HVAE combines variational auto-encoder(VAE) with hyperbolic geometry.
% and achieve competitive performance. 
Last, HGCF applies the latest Hyperbolic GCN \cite{hgcn} to recommendation systems.
%and achieves SOTA performance among various datasets. 

\textbf{Implementation.} For a fair comparison, the embedding dimensionality is set to 50 for all methods, and the same negative sampling strategy is adopted. 
For all baseline methods, suggested settings in original papers are followed. 
Grid search for hyper-parameters are conducted following HGCF \cite{hgcf}.
For LGCF, the number of GCN layers is set to 3. 
We set the learning rate to 0.001 and weight decay to 0.005. And model is trained for 1000 epochs. 
Hyper-parameter margin $m$ is tuned from $[0.01, 2]$.
%We implement LGCF and all baseline methods based on PyTorch \cite{torch}  and run all experiments on a single NVIDIA GeForce RTX 2080 Ti 12GB. 

\subsection{Overall Results}

\begin{table}[]
\caption{Recall results for all datasets.}
\label{tbl:recall}
\resizebox{\linewidth}{!}{
\begin{tabular}{l|c|c|cc|cc|c}
\toprule
Datasets                     & Metrics & BPRMF     & NGCF   & LightGCN & HVAE   & HGCF   & LGCF   \\
\midrule
\multirow{2}{*}{Amazon-CD}   & R@10    & 0.0779 & 0.0758 & 0.0929   & 0.0781 & \underline{0.0962} & \textbf{0.0996} \\
                             & R@20    & 0.1200 & 0.1150 & 0.1404   & 0.1147 & \underline{0.1455} & \textbf{0.1503} \\
\midrule
\multirow{2}{*}{Amazon-Book} & R@10    & 0.0611 & 0.0658 & 0.0799   & 0.0774 & \underline{0.0867} & \textbf{0.0899} \\
                             & R@20    & 0.0794 & 0.1050 & 0.1248   & 0.1125 & \underline{0.1318} & \textbf{0.1360} \\
\midrule
\multirow{2}{*}{Yelp2020}    & R@10    & 0.0325 & 0.0458 & 0.0522   & 0.0421 & \underline{0.0543} & \textbf{0.0573} \\
                             & R@20    & 0.0556 & 0.0764 & 0.0866   & 0.0691 & \underline{0.0884} & \textbf{0.0946}\\
\bottomrule
\end{tabular}
}
\end{table}

\begin{table}[]
\caption{NDCG results for all datasets.}
\label{tbl:ndcg}
\resizebox{\linewidth}{!}{
\begin{tabular}{l|c|c|cc|cc|c}
\toprule
Datasets                     & Metrics & BPRMF     & NGCF   & LightGCN & HVAE   & HGCF   & LGCF   \\
\midrule
\multirow{2}{*}{Amazon-CD}   & N@10    & 0.0610 & 0.0591 & 0.0726   & 0.0629 & \underline{0.0751} & \textbf{0.0780} \\
                             & N@20    & 0.0974 & 0.0718 & 0.0881   & 0.0749 & \underline{0.0909} & \textbf{0.0945} \\
\midrule
\multirow{2}{*}{Amazon-Book} & N@10    & 0.0594 & 0.0655 & 0.0780   & 0.0778 & \underline{0.0869} & \textbf{0.0906} \\
                             & N@20    & 0.0971 & 0.0791 & 0.0938   & 0.0901 & \underline{0.1022} & \textbf{0.1063} \\
\midrule
\multirow{2}{*}{Yelp2020}    & N@10    & 0.0283 & 0.0405 & \underline{0.0461}   & 0.0371 & 0.0458 & \textbf{0.0485} \\
                             & N@20    & 0.0512 & 0.0513 & 0.0582   & 0.0465 & \underline{0.0585} & \textbf{0.0612}   \\
\bottomrule 
\end{tabular}
}
\end{table}

Recall and NDCG results for all datasets are reported in Table \ref{tbl:recall} and Table \ref{tbl:ndcg} respectively. 
We can see that LGCF consistently outperforms other methods on all the three datasets.
Compared with LightGCN, LGCF achieves an improvement up to \textbf{16.15\%} in NDCG@10 on Amazon-Book dataset, demonstrating the superiority of hyperbolic space over Euclidean space in modeling real-world user-item interactions.
  
Among the baseline methods, HGCF is the most competitive counterpart of LGCF.
Even though HGCF adopts hyperbolic space, it resorts to tangent space to conduct aggregation operations, which brings inevitable distortion. 
In contrast, LGCF performs all graph convolution operations in hyperbolic space.
%LGCF obtains a \textbf{7.01\%} improvement of Recall@20 for yelp2020, validating noise 
LGCF outperforms HGCF with wide margins on all the datasets, showing the information loss introduced by tangent space in HGCF.

%\subsection{The Effect of Layer Combination}
%\begin{table}
%\centering
%\caption{Recall@20 results on Amazon-CD dataset for LGCF and no-layer-combination variant with different number of layers from 1 to 4.}
%\label{tbl:layer_comb}
%\begin{tabular}{c|cccc}
%\toprule
%\# layers                  & 1      & 2      & 3      & 4      \\
%\midrule
%LGCF-no-combine       & 0.1369 & 0.1432 & 0.1335 & 0.1269 \\
%LGCF & 0.1369 & 0.1440 & 0.1503 & 0.1456\\
%\midrule
%Rela. Impro. & 0\% & 0.56\% & 12.58\% & 14.73\% \\
%\bottomrule
%\end{tabular}
%\end{table}
%To validate the effect of layer combination strategy in LGCF, we compare LGCF with its no-layer-combination variant with different number of layers from 1 to 4.
%And we report  Recall@20 results on Amazon-CD dataset in Table \ref{tbl:layer_comb}. 
%As we can see, the performance improvement due to layer combination as the number of layer increases.

\subsection{Ablation Study}

\begin{table}[]
\centering
\caption{Results for LGCF and LGCF-tangent on Yelp2020.}
\label{tbl:ablation}
\begin{tabular}{l|llll}
\toprule
             & R@10 & R@20 & N@10 & N@20 \\
             \midrule
LGCF         & 0.0573    & 0.0946    & 0.0485  & 0.0612  \\
LGCF-tangent & 0.0545    & 0.0895    & 0.0463  & 0.0586 \\
\bottomrule
\end{tabular}
\end{table}

%To validate the superiority of our fully-hyperbolic graph convolution over resorting to tangent space, we conduct an ablation study on Yelp2020, the largest one among the three datasets.
To further analyze the effect of fully hyperbolic graph convolution network, we conduct an ablation study on Yelp2020, the largest one among the three datasets.
Since simply replacing fully hyperbolic graph convolution with regular Euclidean graph convolution would drive user/item embeddings out of hyperbolic space, 
we conduct convolution operations in tangent space, and name this model variant as LGCF-tangent.
%We change the fully-hyperbolic graph convolution to tangent space convolution similar to HGCN \cite{hgcn}. 
%Simply replacing fully-hyperbolic graph convolution with regular Euclidean graph convolution would make user/item embeddings out of hyperbolic space.
%Hence, we conduct convolution operations in tangent space, and name this model variant as LGCF-tangent.
% which can be seen as an application of HGCN \cite{hgcn} to recommendation.
%This model variant is named LGCF-tangent.
%Experimental results is shown in Table \ref{tbl:ablation}. 
%As illustrated in Table \ref{tbl:ablation},  wide margin between LGCF and LGCF-tangent demonstrates t
From experimental results shown in Table \ref{tbl:ablation}, we can observe that there is a wide margin between the performance of LGCF and LGCF-tangent.
This is because, in LGCF-tangent, errors caused by tangent space approximation accumulate and spread to the whole graph.
As a result, influence from neighbors, especially high-order neighbors, cannot be captured accurately. 

\subsection{Effect of Embedding Dimensionality}
\begin{figure}
	\centering
	\begin{subfigure}[t]{.48\linewidth}
			\centering
			\includegraphics[width = \linewidth]{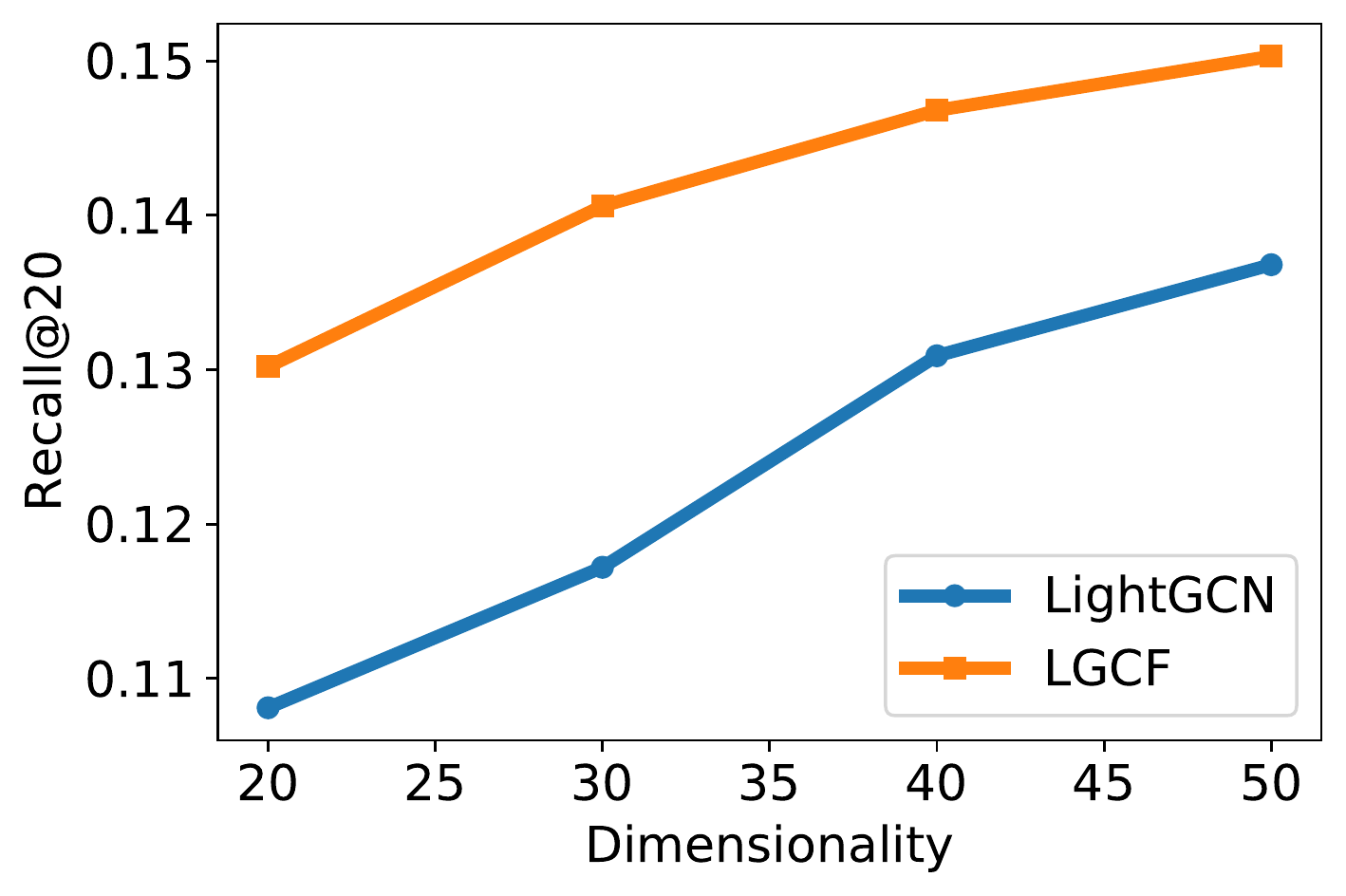} %
%			\caption{}
	\label{fig:single}
	\end{subfigure}
	\begin{subfigure}[t]{.48\linewidth}
			\centering
			\includegraphics[width = \linewidth]{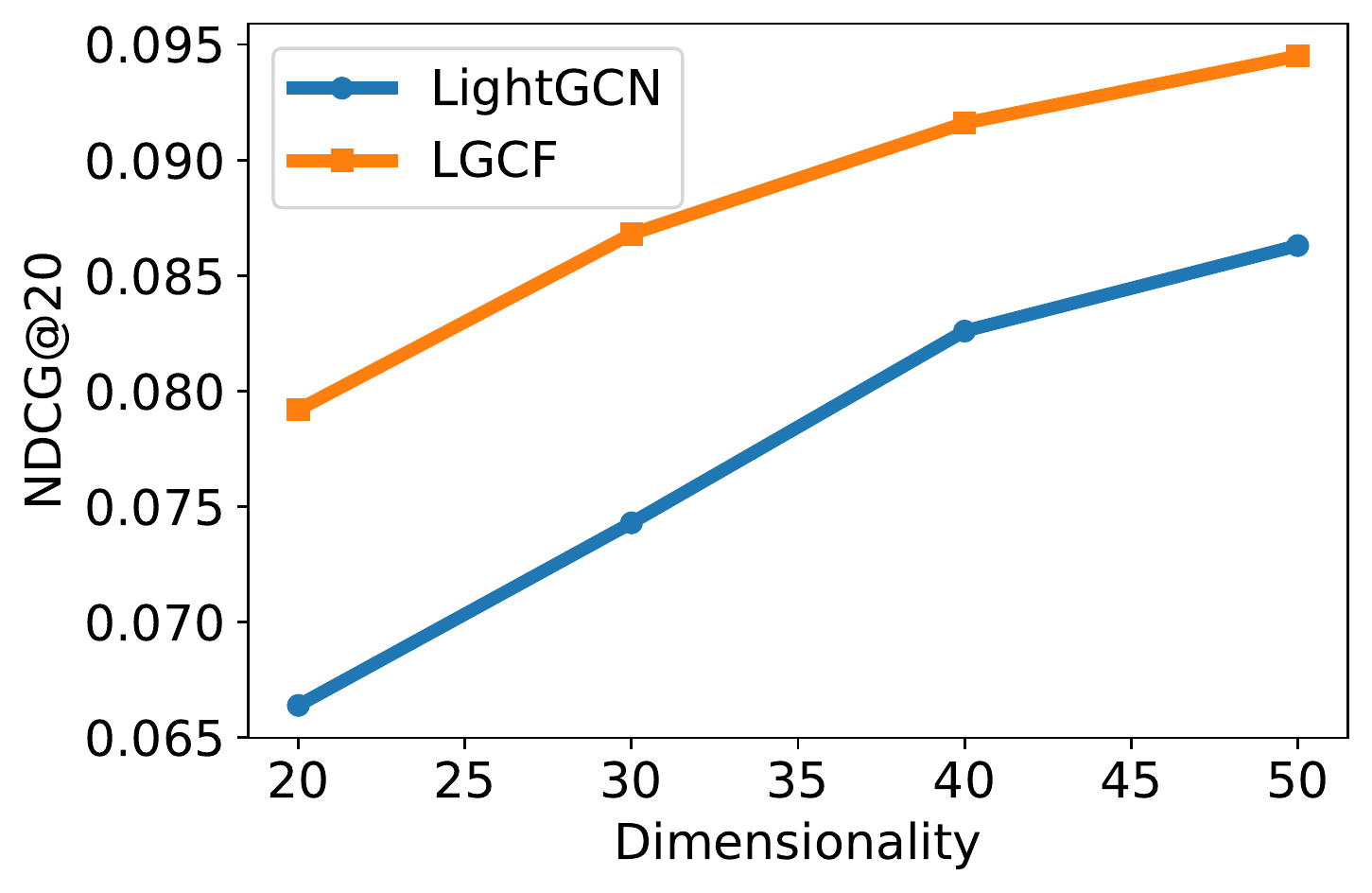} %
%			\caption{}
	\label{fig:single}
	\end{subfigure}
	\caption{Recall@20 and NDCG@20 on Amazon-CD dataset with an embedding dimensionality varying from 20 to 50.}
	\label{fig:dim}
\end{figure}
In order to validate the advantage of hyperbolic space to learn compact representations, 
%we decrease the embedding dimensionality. 
%we show the results of different methods with different embedding dimensionalities.
we compare the results of LGCF and LightGCN with different values of embedding dimensionality.
From Figure \ref{fig:dim}, we can observe that LGCF outperforms LightGCN consistently at all dimensionality values, and the greatest margin occurs at lower dimensionality. 
%reflecting the superiority of hyperbolic space.
%Further, as embedding dimension decreases,  performance improvement of LGCF over LightGCN becomes larger.
LGCF requires far lower embedding dimensionality to achieve comparable performance to its Euclidean analogue.
This reflects that LGCF's advantage is more prominent when the embedding dimensionality cannot be large due to limited computing and storage resources.

%Similar pattern shows in the comparison between LGCF and HGCF, demonstrating the noise introduced by utilizing approximation of tangent space. 
%This phenomenon also proves the advantage of LGCF since all operations are performed in hyperbolic space.
%Experimental results also show that LGCF's advantage is more prominent when the embedding dimensionality can not be large due to limited computing and storage resources.

%% file: conclusion.tex
\section{Conclusion}
In this paper, we propose LGCF, a fully hyperbolic GCN model for recommendation.
%where all operations are performed in the Lorentz model of hyperbolic space.
%LGCF aggregates information 
%Both users and items are embedded in hyperbolic space and they aggregate information from neighbors iteratively.
Utilizing the advantage of hyperbolic space, LGCF is able to embed users/items with less distortion and capture user-item interaction relationship more accurately.
%Extensive experiments on public benchmark datasets show that LGCF outperforms all the existing methods and requires far lower embedding dimensionality to achieve comparable performance to its Euclidean counterparts.
Extensive experiments on public benchmark datasets show that LGCF outperforms both Euclidean and hyperbolic counterparts and requires far lower embedding dimensionality to achieve comparable performance.